\documentclass[aps,prd,preprint,showpacs,groupedaddress,nofootinbib]{revtex4}
\pdfoutput=1
\usepackage{amssymb,amsbsy,color}
\usepackage{latexsym}
\usepackage{graphics}
\usepackage{graphicx}
\usepackage{psfrag}

\newcommand{\be}{\begin{equation}}
\newcommand{\ee}{\end{equation}}
\newcommand{\bea}{\begin{eqnarray}}
\newcommand{\eea}{\end{eqnarray}}
\newcommand{\ba}{\begin{array}}
\newcommand{\ea}{\end{array}}

\newcommand{\pref}[1]{(\ref{#1})}

\def\bqry{\begin{eqnarray}}
\def\eqry{\end{eqnarray}}

\def\lbl{\label}

\newcommand{\Tr}[1]{\langle {#1} \rangle}

\newcommand{\tmu}{\tilde{\mu}}
\newcommand{\muref}{\mu_{\rm eff}}
\newcommand{\tmueff}{\tilde{\mu}_{\rm eff}}
\newcommand{\nn}{\nonumber}

\begin{document}

\preprint{HU-EP/08/60, IFIC/08-63, SFB/CPP-08-102}
\title{
The epsilon regime with Wilson fermions  
}
\author{$^{1}$Oliver B\"ar, $^{2}$Silvia Necco and $^{1}$Stefan Schaefer }

\affiliation{
$^1$Institute of Physics, Humboldt University Berlin, Newtonstrasse
15, 12489 Berlin, Germany\\
$^2$Instituto de F\'isica Corpuscular, CSIC-Universitat de Val\`encia\\
Apartado de Correos 22085, E-46071 Valencia, Spain
}

%
\begin{abstract}
%
We study the impact of explicit chiral symmetry breaking of Wilson fermions on
mesonic correlators in the $\epsilon$-regime  using Wilson chiral perturbation
theory (WChPT). We generalize the $\epsilon$-expansion of continuum ChPT to
nonzero lattice spacings for various quark mass regimes. It turns out that the
corrections due to a nonzero lattice spacing are highly suppressed for typical
quark masses of the order $a\Lambda_{\rm QCD}^2$. The lattice spacing effects
become more pronounced for smaller quark masses and lead to non-trivial
corrections of the continuum ChPT results at next-to-leading order. We compute
these corrections for the standard current and density correlation functions. A
fit to lattice data shows that these corrections are small, as expected.

\end{abstract}
\pacs{11.15.Ha, 12.39.Fe, 12.38.Gc}
\maketitle

\section{Introduction}
\label{sect:intro}

The $\epsilon$-regime of QCD \cite{Gasser:1986vb,Gasser:1987ah} offers various
advantages for the numerical determination of the low-energy couplings (LECs)
in chiral perturbation theory (ChPT), the low-energy effective theory of QCD.
Only the leading order LECs, the pseudo scalar decay constant in the chiral
limit $F$  and the chiral condensate $\Sigma$, enter the predictions of
ChPT through next-to-leading order (NLO) in the epsilon expansion. The
Gasser-Leutwyler coefficients \cite{Gasser:1983yg,Gasser:1984gg} first appear
at one order higher, thus making the $\epsilon$-regime attractive for precise
determinations of $F$ and $\Sigma$.\footnote{For a recent review of the various
estimates see \cite{Necco:2007pr}.} In addition, gauge field topology plays an
important role in the $\epsilon$-regime \cite{Leutwyler:1992yt}. ChPT makes
predictions for correlation functions restricted to individual topological
sectors, thus enlarging the number of observables that can be compared to
numerical lattice QCD results. 

The role of topology and spontaneous chiral symmetry breaking has led to the
widespread conviction that overlap \cite{Neuberger:1997fp} or domain-wall
fermions \cite{Kaplan:1992bt,Shamir:1993zy,Furman:1994ky} are the preferred or
even mandatory choice for lattice simulations in the $\epsilon$-regime.
Consequently, a fairly large number of quenched simulations with these fermions
in the $\epsilon$-regime can be found in the literature \cite{Hernandez:1999cu,DeGrand:2001ie,Hasenfratz:2002rp,Bietenholz:2003bj,Giusti:2004yp,Fukaya:2005yg,Bietenholz:2006fj,Giusti:2007cn,Giusti:2008fz}. Even though the
results were to a large extend promising, the main hurdle for progress in real
QCD is the need for  simulations with dynamical sea quarks, and these are extremely
time-consuming.\footnote{For a recent review see \cite{Schaefer:2006bk}.} So
far only the JLQCD collaboration \cite{Fukaya:2007pn} has carried  out a large
scale dynamical simulation with overlap fermions in the $\epsilon$-regime, and
the computer resources that went into this simulation are enormous.

In contrast, fairly inexpensive simulations with tree-level improved Wilson
fermions have been reported recently \cite{Hasenfratz:2008ce}. Reweighting as
described in Ref.~\cite{Hasenfratz:2008fg} has been used to reach small enough quark
masses in order to be in the $\epsilon$-regime. The size of the box was $L\simeq
2.8$fm, much larger than in all the simulations mentioned before. Quite
surprisingly, the data for the axial vector and pseudo scalar correlation
functions are very well described by the corresponding ChPT predictions,
although chiral symmetry is explicitly broken for Wilson fermions.

A similar observation has been made before by the ETM collaboration
\cite{Jansen:2007rx,Jansen:2008ru}. Their data\footnote{A review of the ETM
results can be found in Ref.\ \cite{Urbach:2007rt}.}, obtained with a twisted
mass term \cite{Frezzotti:2000nk,Frezzotti:2001ea}, also suggests that the
$\epsilon$-regime can be reached with Wilson fermions. One might argue that
automatic O($a$) improvement at maximal twist
\cite{Frezzotti:2003ni,Aoki:2004ta,Aoki:2006nv} may suppress chiral symmetry
breaking effects. Still, the data obtained with Wilson fermions raises the
question if and how the results can be interpreted in the presence of
the explicit chiral symmetry breaking by the Wilson term.

In this paper we study this question with Wilson ChPT
\cite{Sharpe:1998xm,Rupak:2002sm}, the low-energy effective theory for lattice
QCD with Wilson fermions. And indeed, our analysis suggests a natural answer to
the question raised above.  It turns out that the lattice spacing corrections
are in general highly suppressed and show up at higher order in the epsilon
expansion. For example, for quark masses $m\sim a\Lambda_{\rm QCD}^2$ the
deviations from the continuum results due to the O($a$) corrections enter first
at next-to-next-to-leading order (NNLO). This is in contrast to the expansion in the $p$-regime, where the
corrections already appear at next-to-leading order. This suppression is completely analogous to
the suppression of terms involving the Gasser-Leutwyler coefficients in
continuum ChPT in the $\epsilon$-regime.

However, the power counting in WChPT can be different, depending on the
relative size of the quark mass $m$ and the lattice spacing $a$. For this
reason we also consider a different power counting where $m\sim a\Lambda_{\rm
QCD}^2$ is no longer appropriate. In this case the lattice spacing corrections
enter already at NLO. Interestingly, these corrections are entirely caused by
the O($a^2$) term in the chiral effective Lagrangian that also determines the
phase diagram of the lattice theory \cite{Sharpe:1998xm}. The corrections
linear in the lattice spacing, stemming from the effective Lagrangian and the
effective operators, are still of higher order in the epsilon expansion. This is
relevant in practice: The Wilson ChPT expressions contain only one more unknown
LEC at this order, and the predictive power is not spoiled by a plethora of
free fit parameters.


We also use our WChPT results derived here for an analysis of the data of Ref.\
\cite{Hasenfratz:2008ce}. The corrections due to the nonzero lattice spacing
turn out to be very small, supporting our theoretical analysis that these
corrections are in general highly suppressed. This is very encouraging for
lattice simulations with Wilson fermions. The impact of explicit chiral
symmetry breaking for $\epsilon$-regime simulations is much less severe than
previously thought. This makes simulations with Wilson fermions a serious and
efficient alternative to those with chiral fermions.

\section{Wilson chiral perturbation theory (WChPT)}
\label{sect:tree}

\subsection{Chiral Lagrangian}
\label{ssect:setup}

The chiral effective Lagrangian of WChPT is expanded in powers of (small) pion
momenta $p^{2}$, quark masses $m$ and the lattice spacing $a$. Based on the
symmetries of the underlying Symanzik action
\cite{Symanzik:1983dc,Symanzik:1983gh} the chiral Lagrangian including all
terms of  O$(p^{4},p^{2}m,m^2,p^{2}a,ma)$ is given in Ref.\
\cite{Rupak:2002sm}. The O$(a^{2})$ contributions are constructed in Ref.\
\cite{Bar:2003mh} and, independently,  in Ref.\ \cite{Aoki:2003yv} for the
two-flavor case.  

In the following we will restrict ourselves to $N_{f}=2$  with degenerate quark
mass $m$.  The continuum part of
the chiral Lagrangian is the well-known Gasser-Leutwyler Lagrangian
\cite{Weinberg:1978kz,Gasser:1983yg}. With our notations the leading part reads
(in Euclidean space-time) 
\bqry
\lbl{L2}
{\cal L}_{2} & =& \frac{F^{2}}{4} {\rm Tr}
\left(\partial_{\mu}U \partial_{\mu}U^{\dagger}\right) -
\frac{F^{2}B}{2} {m} {\rm Tr}\left( U + U^{\dagger}\right). 
\eqry
The field $U$ containing the pion fields is defined as usual, 
\bqry
\lbl{Sigma}
U(x) &=& \exp \left( \frac{2 i}{F} \xi(x)\right), \qquad \xi(x)\,=\, \xi^{a}(x) T^a.
\eqry
The SU(2) generators are normalized such that ${\rm Tr}
(T^aT^b)=\delta^{ab}/2$, so $T^a=\sigma^a/2$ in terms of the standard Pauli
matrices $\sigma^a$. The coefficients $B$ and $F$ are the familiar leading
order (LO) low-energy coefficients.\footnote{With our normalization the
pion decay constant in the chiral limit is $F\approx 93$MeV.}
Higher order terms are collected in the next-to-leading order Lagrangian ${\cal
L}_4$ \cite{Gasser:1983yg}, which we do not need in this work.

The terms involving the lattice spacing are as follows:\footnote{We essentially
adopt the notation of Refs.\ \cite{Sharpe:2004ny} and \cite{Aoki:2006nv}.} 
\bea
{\cal L}_a&=& \hat{a} W_{45} {\rm Tr}\left( \partial_{\mu}U \partial_{\mu}U^{\dagger}\right)  
{\rm Tr}\left( U +U^{\dagger}\right) - \hat{a}\hat{m} W_{68}   ({\rm Tr}\left( U +U^{\dagger}\right))^2\,,\label{La}\\
{\cal L}_{a^2}&=& \frac{F^{2}}{16} c_2a^2( {\rm Tr}\left({U+U^\dagger}\right))^2.\label{Lasquare}
\eea
$W_{45},W_{68}$ are LECs, similar to Gasser-Leutwyler coefficients in ${\cal
L}_4$. The quark mass and lattice spacing enter through the combinations
\bqry
\lbl{qmass}
\hat{m} & = & 2Bm\,,\qquad \hat{a} \, =\, 2W_0 a\,,
\eqry
where $W_{0}$ is another low-energy coefficient \cite{Rupak:2002sm}. Its
presence here is for dimensional reasons: $W_{0}$ is of dimension three and
$\hat{a}$ therefore of dimension two. Hence, $\hat{a}$ has the same dimension
as the familiar combination $Bm$.

We have chosen to parametrize the O($a^2$) contribution in terms of the LEC
$c_2$. This coefficient plays a prominent role since its sign determines the
phase diagram of the theory \cite{Sharpe:1998xm}.  We briefly come back to this
after we have discussed the power counting in WChPT in section \ref{ssect:pc}.

Note that the mass parameter $m$ in eq.\ \pref{L2} is the so-called {\em
shifted mass} \cite{Sharpe:1998xm}. Besides the dominant additive mass
renormalization proportional to 1/$a$ it also contains the leading correction
of O($a$). Consequently, the term $F^2\hat{a}{\rm
Tr}\left(U+U^{\dagger}\right)/4$ is not explicitly present in the chiral
Lagrangian since it is absorbed in the shifted mass \cite{Rupak:2002sm}. Of course, physical
results expressed by observables do not depend on what mass is used for the
parametrization of the chiral Lagrangian. 

\subsection{Currents and densities}
\label{ssect:CurrDen}

The expressions for the currents and densities in continuum ChPT are well
known. For example, the LO expressions for the axial vector current and the
pseudo scalar density read 
\bea
A_{\mu,{\rm ct}}^{a} & = & i\frac{F^{2}}{2} {\rm Tr}\left({T^{a}( U^{\dagger}\partial_{\mu}U-U\partial_{\mu}U^{\dagger})}\right)\,,\label{Axialcurrentct}\\
P^{a}_{\rm ct} & = & i\frac{F^{2}B}{2} {\rm Tr}\left({T^{a}(U - U^{\dagger})}\right)\,.\label{Pcont}
\eea
In WChPT these expressions receive corrections proportional to the lattice
spacing. The currents and densities in WChPT can be constructed by a standard
spurion analysis, similarly to the construction of the chiral Lagrangian. One
first writes down the most general current/density that is compatible with the
symmetries using the chiral field $U$, its derivatives and the spurion fields.
Then, in a second step, one imposes the appropriate Ward identities valid in
the theory. For the vector and axial vector current this has been done in
\cite{Aoki:2007es}. Alternatively one can introduce source terms for the
currents and densities and constructs the generating functional, as has been
done in Ref.\ \cite{Sharpe:2004ny}. Carrying over the notation of this
reference the axial vector current including the leading corrections of O($a$)
reads 
\bea\label{AVWChPT}
A_{\mu,{\rm WChPT}}^{a} & = & A_{\mu,{\rm cont}}^{a} \left(1 +
  \frac{4}{F^{2}}\hat{a}{W}_{45}{\rm Tr}\left(U + {U^{\dagger}}
  \right)\right) + 2\hat{a}{W}_{10}\partial_{\mu} {\rm Tr}\left({T^{a}(U - U^{\dagger})}\right)\,.
\eea

This axial vector does not satisfy any specific
renormalization condition. Imposing a particular renormalization condition
leads to a finite renormalization. Explicitly, one introduces
\cite{Aoki:2007es} 
\bea\label{ZAeff}
A_{\mu,{\rm ren}}^{a}(x) & = & Z_{A} A_{\mu,{\rm WChPT}}^{a}(x)
\eea
with a renormalization factor $Z_A$.\footnote{$Z_A$ is a renormalization factor
in the effective theory and should not be confused with $Z_A$ in the underlying
lattice theory.} Now one can  impose a  renormalization condition and
demands that it is satisfied by $A_{\mu,{\rm ren}}^{a}$; this determines $Z_A$.
For instance, in Ref.\ \cite{Aoki:2007es} the massless chiral Ward identity has
been imposed.

Quite generally, $Z_A$ has the form (up to O($a$))
\bea\label{DefZAGen}
Z_A&=& 1 + \frac{16}{F^2}\hat{a} W_{A}
\eea
with an unknown coefficient $W_{A}$. This form reflects the fact that, by
construction, the WChPT current reduces to the correct continuum current for
$a\rightarrow 0$. Consequently, $Z_A$ is equal to one in the continuum limit,
and \pref{DefZAGen} is the leading generalization for $a\neq0$. Hence, the
general form of the renormalized current is, up to O($a$),
\bea\label{AVWChPT2}
A_{\mu,{\rm WChPT}}^{a} & = & A_{\mu,{\rm cont}}^{a} \left(1 +
  \frac{4}{F^{2}}\hat{a}\left[{W}_{45}{\rm Tr}\left(U + U^{\dagger} 
 \right)+4W_{A}\right] \right)\nn\\
& & \,\,\,+
2\hat{a}{W}_{10}\partial_{\mu} {\rm Tr}\left({T^{a}(U - U^{\dagger})}\right).
\eea
For brevity we have dropped the subscript ``ren" on the left hand side.
Terms of O$(am,a^2)$ will be present at higher order in the chiral expansion.

Analogously, we use the expression \cite{Sharpe:2004ny} 
\bea\label{eq:Pseudoscalareff}
P^{a}_{\rm WChPT} & = & P^{a}_{\rm Cont}\left(1 +
  \frac{4}{F^{2}}\hat{a}\left[{W}_{68} {\rm Tr}\left({U^{\dagger} + U}\right) + 4W_{P}\right]\right)
\eea
for the pseudo scalar density. The contribution involving the LEC $W_{P}$ (not
present in Ref.\ \cite{Sharpe:2004ny}) stems from a general renormalization
factor $Z_P =1+16\hat{a} W_{P}/F^2$, which we also allow, even though the
results derived in this paper will not depend on the details of  the O($a$)
correction.

\subsection{Power counting in infinite volume}
\label{ssect:pc}

In WChPT there are two parameters that break chiral symmetry explicitly, the
quark mass $m$ and the lattice spacing $a$.  The power counting is determined
by the relative size of these two parameters. 

The literature \cite{Sharpe:2004ps,Sharpe:2004ny} distinguishes two  quark mass
regimes with different power countings: (i) the GSM regime\footnote{GSM stands
for {\em generically small masses}.} where $m\sim a\Lambda_{\rm QCD}^2$  and
(ii) the Aoki regime where  $m\sim a^2\Lambda_{\rm QCD}^3$. A priori one does
not know in which regime one actually has performed a simulation. For this to
decide one has to compare with the predictions of WChPT and check which
expressions fit the data better. However, recalling how lattice simulations are
typically done one can easily imagine that one starts in the GSM regime and by
lowering the quark mass at fixed lattice spacing one will eventually enter the
Aoki regime. 

Depending on the particular regime, the LO Lagrangian is different. Since
$m\sim a^2\Lambda_{\rm QCD}^3$ in the Aoki regime, also the ${\cal L}_{a^2}$
part in \pref{Lasquare} counts as LO \cite{Aoki:2003yv}:
\bea
{\rm GSM\, regime:}& \quad &{\cal L}_{\rm LO} \,=\, {\cal L}_{2} \,,\nonumber\\
{\rm Aoki\, regime:}&\quad& {\cal L}_{\rm LO} \,=\, {\cal L}_{2} + {\cal L}_{a^2} \,.\nonumber
\eea
The effects due to a nonzero lattice spacing are much more pronounced in the
Aoki regime. Non-trivial phase transitions become relevant \cite{Sharpe:1998xm}
and additional chiral logarithms proportional to $a^2$ appear at one loop
\cite{Aoki:2003yv,Aoki:2008gy}. 

\subsection{The pion mass and the PCAC mass in infinite volume}
\label{ssect:pregime}

It is  useful to derive the pion mass and PCAC mass at LO.\\ 
We start with the calculation of the pion mass. Expanding the LO chiral
Lagrangian to quadratic order in the pion fields we obtain
\bea
{\rm GSM \,regime:}\qquad M_{0}^2 & = & 2Bm\,,\label{MpiGSM}\\
{\rm Aoki\, regime:}\qquad M_{0}^2 & = & 2Bm - 2c_2a^2\label{MpiAoki}\,.
\eea
The sign of $c_2$ determines the phase diagram of the theory
\cite{Sharpe:1998xm}.\footnote{Note that our definition for $c_2$ differs by a
factor $F^2a^2$ from the one in Ref.\ \cite{Sharpe:1998xm}.} For $c_2>0$ there
exists a second-order phase transition separating the Aoki phase
\cite{Aoki:1983qi}. The charged pions are massless in this phase due to the
spontaneous breaking of the flavor symmetry. The pion mass vanishes at $m =
c_2a^2/B$. For even smaller values of $m$ the charged pions remain massless,
while the neutral pion becomes massive again \cite{Sharpe:1998xm}.

Negative values of $c_2$, on the other hand, imply a first order phase
transition with a minimal non-vanishing pion mass. All three pions are massive
for all quark masses, and the pion mass assumes its minimal value at $m=0$,
resulting in 
\bqry
\lbl{Mpimin}
M_{0,{\rm min}}^2 & =& 2|c_2|a^2\,.
\eqry

Note that magnitude and the sign of $c_2$ are a priori unknown and depend on
the details of the underlying lattice theory, i.e.\ what particular lattice
action has been used. 

The PCAC quark mass is defined by the ratio (no sum over $a$)
\bea\label{DefmPCAC}
m_{\rm PCAC} &=& \frac{\langle \partial_{\mu} A_{\mu}^a(x) P^a(0)\rangle}{2\langle P^a(x) P^a(0)\rangle}\,,
\eea
where angled brackets indicate expectation values.
Expanding the current and the density in Eqs.~\pref{AVWChPT2} and
\pref{eq:Pseudoscalareff} to O($\xi^a$) we find 
\bea
A_{\mu}^a(x)&=&  -iF\partial_{\mu}\xi^a(x)\left(1+ a c_A\right)\,,\\
P^a(x) &=& iFB\xi^a(x)\left(1+ a c_P\right)\,.
\eea
where here and in the following we for simplicity no longer write subscript ``WChPT''.
We also  introduced the short hand notation 
\bea
c_A &=& \frac{16}{F^2}2W_0 [ W_{45} + W_{A} - W_{10}/4]\,, \\
c_P &=&  \frac{16}{F^2}2W_0 [ W_{68} + W_{P}]\,,
\eea
for the combinations of LECs in the effective current and density at this order.
The correlation functions in \pref{DefmPCAC} are now easily computed at LO, yielding
\bea
m_{\rm PCAC} &=& \frac{M_{0}^2}{2B}\Big(1 + a(c_A - c_P)\Big)\;.
\eea
Using the tree-level pion mass obtained above we find
\bea
{\rm GSM \,regime:}&\quad& m_{\rm PCAC} \, = \,m\Big(1 + a(c_A - c_P)\Big)\,,\label{mpcacGSM}\\
{\rm Aoki\, regime:}&\quad& m_{\rm PCAC} \, = \, \left(m - \frac{c_2}{B}a^2\right)\Big(1 + a(c_A - c_P)\Big)\label{mpcacAoki}\,.
\eea
In the GSM regime the result is rather simple and the PCAC mass is equal to the
shifted quark mass, up to corrections of O($ma$). This is no longer true in the
Aoki regime. Still, the contribution proportional to $(c_A - c_P)$ is
subleading and can be ignored if one works to leading order in the quark mass.

Equations \pref{mpcacGSM} and \pref{mpcacAoki} allow to replace the shifted
mass $m$, which is just a parameter in the chiral Lagrangian, with the PCAC
quark mass. The latter is an observable which is often used in lattice
simulations. 

Note that the results above reproduce  a well-known fact, namely that the PCAC
mass depends on the particular renormalization conditions imposed on the axial
vector current and the pseudo scalar density. Different renormalization
conditions show up as different values of $c_A$ and $c_P$.

\section{WChPT in the epsilon regime}
\label{sect:epsWChPT}

\subsection{Continuum ChPT in finite volume}
\label{ssect:PCcont}

Consider continuum QCD with $N_f$ degenerate quark masses in a hypercubic volume $V=TL^3$, with $T,L\gg 1/\Lambda_{\rm
QCD}$. Finite-size effects can be systematically studied by means of 
the corresponding chiral effective theory \cite{Gasser:1986vb,Gasser:1987ah,Gasser:1987zq}. 
In this section we
summarize the main aspects of finite-volume chiral perturbation theory in the continuum.

If the pion Compton wavelength is much smaller than the size of the
box, $M_\pi L\gg 1$, finite-volume effects can be treated in the
chiral effective theory by adopting the standard $p$-expansion, where
the power-counting in terms of the momentum $p$ 
is given by 
\begin{equation}
m\;\sim O(p^2), \;\;\;\;1/L,\;1/T\;,\partial_\mu\; \sim O(p),\;\;\;\; \xi\;\sim O(p).
\end{equation}
For asymptotically large volumes, one expects the finite-volume
effects to be exponentially suppressed by factors $\sim e^{-M_\pi L}$. 

On the other hand, approaching the chiral limit by keeping
$\mu=m\Sigma V\lesssim O(1)$ (but still $L\gg 1/\Lambda_{\rm QCD}$),
where  $\Sigma=F^2B$ is the quark condensate in the chiral limit,
one explores the domain where the pion wavelength is larger than the
size of the box, $M_\pi L<1$. 
In this case the pion zero-mode gives a
contribution to the propagator proportional to $1/M_0^2V$,
which cannot be treated perturbatively but has to be
computed exactly 
\cite{Gasser:1986vb,Gasser:1987ah}. This is achieved by
factorizing the pseudo Nambu-Goldstone boson fields as
\bea\label{eps_fac}
U(x) &=& \exp\left(\frac{2i}{F}\xi(x)\right)U_0\,,
\eea
where the constant $U_0\in SU(N_f)$ represents the collective zero-mode. The nonzero modes parametrized by $\xi$, on the
other hand, can still be treated perturbatively and satisfy the condition
\bea\label{nonconstmode}
\int_V d^4 x\, \xi(x)&=&0,
\eea
since the constant mode has been separated. 

The zero-mode contribution proportional to $1/M_0^2 V$ in the pion propagator
diverges in the chiral limit and a reordering of the perturbation series that
sums all graphs with an arbitrary number of zero-mode propagators is necessary
\cite{Gasser:1987ah}. 
This reordering is achieved 
with the power counting
\begin{equation}
m\;\sim O(\epsilon^4), \;\;\;\;1/L,\;1/T\;,\partial_\mu\; \sim O(\epsilon),\;\;\;\; \xi\;\sim O(\epsilon).
\end{equation}
Mass effects are suppressed compared to the $p$-regime, while
volume effects are enhanced and become polynomial in $(FL)^{-2}$.
Since $M_0^2$ is proportional to $m$ the combination $1/M_0^2 V$ now counts as $\epsilon^0$. 
Consequently, all graphs that exclusively involve zero-mode propagators count as O$(1)$ and are unsuppressed. 
The key point here is, that the counting of the quark mass is dictated by the
counting of $L$ by demanding $1/M_0^2 V = {\rm O}(\epsilon^0)$. We will use
this in the next section in order to establish the counting rules in WChPT.

With the factorization given in Eq. \pref{eps_fac}, the leading order continuum partition function in the $\epsilon$-regime is given
by 
\bea\label{zetaep}
Z=\int_{SU(N_f)}[dU_0]\int [d\xi] \exp\left\{\frac{1}{2}\int_V d^4 x {\rm
    Tr}(\partial_\mu\xi\partial_\mu\xi) +\frac{m\Sigma V}{2}{\rm Tr}(U_0+U_0^\dagger)\right\}.
\eea
The integration over the perturbative degrees of freedom
$[d\xi]$ gives
rise to the usual Wick contractions, while the zero-mode integrals
over $[dU_0]$ must be computed exactly.
Notice that by going to $O(\epsilon ^2)$, by factoring out the constant zero-mode from the measure, one obtains
\bea
[dU]=[d\xi][dU_0]\left(1+A(\xi)+O(\epsilon^4)\right),
\eea
with
\bea
A(\xi)=-\frac{2N_f}{3F^2}\frac{1}{V}\int_V d^4 x\, {\rm Tr}(\xi^2(x))
\eea
for a general value of $N_f$ \cite{Gasser:1987ah,Hansen:1990un}.

\subsection{Power countings for the epsilon regime in WChPT}
\label{ssect:PC2}
Like the continuum effective theory, WChPT can also be  formulated in a finite volume,
in particular the $\epsilon$-regime discussed in this section.\\
In WChPT we have additional low-energy constants and the lattice spacing as an
additional expansion parameter. The main task is to decide how to count these
in the epsilon expansion. 

Just as the continuum LECs $F$ and $\Sigma$, we count all the additional LECs
associated with the lattice spacing to be of order $\epsilon^0$,
\bea
&& c_2, c_A, c_P  \sim{\rm O}(1).  
\eea
The counting of the lattice spacing $a$ is more complicated. The
general strategy is to follow the infinite-volume procedure and
determine the power counting depending on the relative size of $m$ and
$a$. At finite volume, once  the counting of $m$ is fixed by the counting of
$L$, we obtain the counting of $a$. 

We start with the GSM regime. The LO Lagrangian and the pion mass
$M_0^2$ are as in the continuum $\epsilon$-regime, so we conclude $m\sim{\rm O}(\epsilon^4)$ 
by the same arguments as in the previous section. Since the GSM
regime is defined by $m\sim a\Lambda_{\rm QCD}^2$ we are immediately lead to
$a\sim 
{\rm O}(\epsilon^4)$. 

The Aoki regime is more subtle. According to our assumption, 
the pion mass $M_0^2$, given in \pref{MpiAoki},
is now a sum of  two terms of equal order. If
$c_2<0$, it is a sum of two positive terms. Hence, a small pion mass of order
O($\epsilon^4$)  implies that both terms, $2Bm$ and $2|c_2|a^2$ are small too
and also of order O($\epsilon^4$). 

If $c_2$ is positive, the pion mass is the difference of two positive
contributions. This leaves the possibility that $M_0^2$ is small, even though
the individual terms $2Bm$ and $2|c_2|a^2$ may not be small and only their
difference is. A pion mass of order $\epsilon^4$ may be obtained by the
difference of two order $\epsilon^2$ or $\epsilon^3$ terms, for example. 

We do not think that this is a likely scenario. Present day lattice simulations
are usually done with small lattice spacings less than 0.1 fm and the O$(a^2)$
corrections are expected to be small in this case. Hence, we {\em assume} that
$a^2\sim {\rm O}(\epsilon^4)$ in the Aoki regime. This assumption, together
with the requirement $M_0^2\sim {\rm O}(\epsilon^4)$ then also leads to
$m\sim{\rm O}(\epsilon^4)$, the same counting as for $c_2<0$.  

The epsilon expansion allows us to introduce yet another regime where we count
$a\sim \epsilon^3$. Just by the powers of $\epsilon$ this is an intermediate
regime between the GSM and Aoki regime. 
One may think about it as the GSM regime but at a larger lattice spacing (or
smaller quark mass).
Its usefulness will become clear in the next section when we discuss the epsilon expansion of correlation functions.

All three countings we introduced are well defined and are appropriate for a
particular relative size between $m$ and $a$. In order to be able to refer to
these regimes we introduce the following nomenclature:
\bea
{\rm GSM\, regime:}\qquad a &\sim & {\rm O}(\epsilon^4)\,,\nonumber\\
{\rm GSM^{\ast}\, regime:}\qquad a &\sim & {\rm O}(\epsilon^3)\,,\\
{\rm Aoki\, regime:}\qquad a &\sim & {\rm O}(\epsilon^2)\,.\nonumber
\eea
For fixed values of $m$ and $a$ in a given regime, one can match for instance the
time-dependence of current correlators with lattice QCD results
in order to extract the corresponding LECs.  

\subsection{Epsilon expansion of correlation functions}
\label{ssect:PC}

We will be interested in correlators of the pseudo scalar density and the axial
vector current. These correlators have been calculated before through NNLO in
continuum ChPT \cite{Hansen:1990un}. In powers of $\epsilon$ this corresponds
to $O(\epsilon^4)$ for the $\langle P^a(x)P^a(0)\rangle $ correlator and
$O(\epsilon^8)$ for $\langle A_{\mu}^a(x)A_{\mu}^a(0)\rangle $.

In order to discuss the epsilon expansion in WChPT let us split an arbitrary
operator and the action in WChPT into the continuum part and a remainder
proportional to powers of $a$,
\bea
 O(x) &=& O_{\rm ct}(x) + \delta O(x)\,,\\
S & = &S_{\rm ct} + \delta S\,.
\eea
Expectation values are generically defined as
\bea
\langle O \rangle =\frac{1}{Z}\int [dU]e^{-S}O,
\eea
where $Z$ is the partition function
\bea
Z=\int [dU]e^{-S}.
\eea
The two-point correlator $\langle O_1(x) O_2(0)\rangle = \langle O_1
O_2\rangle$ (for notational simplicity we suppress the dependence on $x$) can
then be written according to
\bea
\langle O_1 O_2\rangle_W & =&   \langle O_{1,{\rm ct}}O_{2,{\rm ct}}\rangle + \delta \langle O_1O_2\rangle\,,\label{Corr}\\
 \delta \langle O_1O_2\rangle &=&   \langle O_{1,{\rm ct}}\delta O_2 + \delta O_1 O_{2,{\rm ct}}\rangle - \langle O_{1,{\rm ct}}O_{2,{\rm ct}}\delta S \rangle + \langle O_{1,{\rm ct}}O_{2,{\rm ct}}\rangle \langle\delta S \rangle\,.\label{deltaCorr}
\eea
Here we have approximated $\exp(-\delta S)\approx 1 - \delta S$ and we dropped
all higher corrections. 
Note that the expectation value on the left hand side of
\pref{Corr}, labelled with a subscript ``W'',  is defined with the full action $S$ in the Boltzmann factor, while
on the right hand side it is defined with $S_{\rm ct}$ only (for notational simplicity we suppress a subscript ``ct'').

The discretization corrections for the pseudo scalar and the axial vector can
be read off from \pref{AVWChPT2} and \pref{eq:Pseudoscalareff}. For what
matters here we can simplify these expressions. We are interested in the power
counting for the epsilon expansion, and for this  we can ignore all constants
which count as O($1$). Therefore, we write 
\bea
\delta P^a & \propto & a ( {\rm Tr}\left( U + U^{\dagger}\right)  + 1) P^a_{\rm ct}\,.
\eea
$\delta P^a$ is proportional to the continuum density itself. As mentioned
before,  the epsilon expansion of $P^a_{\rm ct}$ starts with O($\epsilon^0$).
The ``scalar density" 
${\rm Tr} (U + U^{\dagger})$ also starts at O($\epsilon^0$). 
Hence, by considering the continuum contribution at LO, 
the leading correction in the epsilon expansion due to lattice terms is
completely determined by how we count the lattice spacing $a$. In the last
section we defined three different countings, so for now we leave it
unspecified, write $a\sim \epsilon^{n_a}$ where $n_a$ counts the epsilon powers
for $a$, and obtain
\bea
\delta P^a &\sim& \epsilon^{n_a}. 
\eea 
Note that the symbol $\sim$ stands here just for the leading lattice contribution in
the epsilon expansion. 

Analogously, we find for the axial vector (dropping again irrelevant constants)
\bea
\delta A_{\mu}^a & \propto & a \big[ ({\rm Tr} \left( U +
  U^{\dagger}\right) + 1) A^a_{\mu, {\rm ct}} + \partial_{\mu} {\rm Tr}\left(T^a (U - U^{\dagger})\right) \big]\,.
\eea
Both, $ A^a_{\mu, {\rm ct}}$ and  $ \partial_{\mu} {\rm Tr}\left( T^a (U -
U^{\dagger}\right)$ have an open Lorentz index and, therefore, contain at least
one derivative acting on at least one power of $\xi(x)$. Hence, their continuum
epsilon expansion starts at O($\epsilon^2$) 
and we find for the leading lattice corrections
\bea
\delta A_{\mu}^a &\sim& \epsilon^{n_a+2}\,.
\eea
Finally, we have to take into account the lattice corrections due to the
contribution $\delta S$. It will be useful to split them into two parts. The
first one, denoted by $\delta S_a$, contains the terms of ${\cal L}_a$ in
\pref{La}.  These terms start at O($\epsilon^{n_a}$) (having taken into account
$\epsilon^{-4}$  from the integration over space-time),
\bea
\delta S_a & \sim & \epsilon^{n_a}.  
\eea 
The second contribution, $\delta S_{a^2}$, contains only the $a^2$ term proportional to $c_2$.  Therefore, it counts as
\bea
\delta S_{a^2} & \sim & \epsilon^{2n_a-4}.
\eea
After these preparations we can  determine at which order the lattice spacing effects enter the PP and the AA correlator.

\subsubsection{GSM regime}
In the GSM regime we set $n_a=4$ and find
\bea
\delta \langle P^a(x)P^a(0)\rangle & \sim & {\rm O}(\epsilon^4)\,,\\
\delta \langle A^a_{\mu}(x)A^a_{\mu}(0)\rangle & \sim & {\rm O}(\epsilon^8)\,.
\eea
The corrections due to the lattice spacing first affect both correlators at
NNLO. Up to NLO the results obtained in continuum ChPT are the appropriate
ones. This is quite remarkable and may explain why numerical data generated
recently \cite{Hasenfratz:2008ce} could be fitted very well using the NLO
continuum expressions. Note that this suppression to NNLO holds for the
unimproved theory. The reason is that the terms linear in $a$ are accompanied
by at least one additional power of $m$ or $\partial_{\mu}\xi$, and are
therefore of higher order. 

\subsubsection{GSM$^{\ast}$ regime}
Here we have $n_a=3$ and obtain
\bea
\delta \langle P^a(x)P^a(0)\rangle & \sim & {\rm O}(\epsilon^2)\,,\\
\delta \langle A^a_{\mu}(x)A^a_{\mu}(0)\rangle & \sim & {\rm O}(\epsilon^6)\,,
\eea
hence the corrections enter at NLO. Interestingly, the dominant term
here comes only from the correction $\delta S_{a^2}$. The other
corrections from the O($a$) contributions in the currents, densities
and $\delta S_a$ start at $\epsilon^3$ and  $\epsilon^7$, respectively. Therefore,  they are of higher
order in the epsilon expansion, 
even though they are still lower than the NNLO contributions, which start at $\epsilon^4$ and $\epsilon^8$, respectively.

Notice that in the GSM and GSM$^{\ast}$ regimes the leading order partition function is
like the continuum one given in Eq. \pref{zetaep}. In particular, the
exact zero-mode integrals are computed with respect to the same  Boltzmann factor as in continuum ChPT.

\subsubsection{Aoki regime}
The modifications in the Aoki regime are more pronounced than in the
previously discussed regimes. Here, cut-off effects show up already at
LO. Even worse, the corrections can no longer be linearly added to the
continuum result. The reason is the correction $\delta S_{a^2}$, which
gives a zero-mode contribution of order $\epsilon^0$. Hence, it is no longer
justified to expand completely the exponential $\exp(-S_{a^2})\approx 1 -
S_{a^2}$. 
The zero-mode contribution of order $\epsilon^0$ has to be included
in the leading order Boltzmann factor. That is, the partition function becomes
\bea\label{zetaepAoki}
Z_{\rm Aoki}=\int_{SU(2)}[dU_0]\int [d\xi]\exp\Bigg\{\frac{1}{2}\int_V d^4 x {\rm
    Tr}(\partial_\mu\xi\partial_\mu\xi) +\frac{m\Sigma V}{2}{\rm
    Tr}(U_0+U_0^\dagger) 
\eea
$$
-\frac{c_2F^2a^2V}{16}({\rm
    Tr}(U_0+U_0^\dagger))^2  \Bigg\}.
$$
This modification affects all constant integrals and probably leads to
non-trivial changes of the continuum results. Note that the other O($a$)
corrections (from $\delta O$ and $\delta S_a$) are of order $\epsilon^2$ and
show up at NLO only.

\subsection{Comment on O($a$) improvement}

The results in the previous section are valid for unimproved Wilson fermions.
It is natural to ask how (non-perturbative) O($a$)--improvement changes these
results. 

If the theory is non-perturbatively improved the corrections $\delta O$ and
$\delta S_a$ are absent, and modifications are caused by $\delta S_{a^2}$ only.
We have seen that this term is the dominant correction and the others are
subleading. Consequently, the epsilon expansion is essentially unaltered for
the improved theory, since only subleading terms vanish.

\section{Leading correction in the GSM$^{\ast}$ regime}

We already mentioned in the introduction that the epsilon expansion is advantageous for the 
determination of $F$ and $\Sigma$, since the Gasser-Leutwyler coefficients first enter the 
ChPT formulae at NNLO. The same is true for WChPT in the GSM regime, where the additional 
lattice spacing contributions enter at NNLO too. In other words, working through NLO the 
results for the GSM regime are the same as those in continuum ChPT. 

The first non-trivial modification of the continuum NLO results appears in the GSM$^{\ast}$ regime, and
in this section we compute the leading correction to the PP and AA
correlator; the results for some other
correlators are given in appendix \ref{ocorr}.
This correction is caused by the constant term of the $\delta
S_{a^2}$ contribution, 
\bea
 \delta \langle O_1(x)O_2(y)\rangle\Big|_{\rm leading} &=&  - \langle O^{\rm LO}_{1,{\rm ct}}(x)O^{\rm LO}_{2,{\rm ct}}(y)\delta S_{a^2} \rangle +  \langle O^{\rm LO}_{1,{\rm ct}}(x)O^{\rm LO}_{2,{\rm ct}}(y)\rangle\langle\delta S_{a^2} \rangle\,, \label{Defcorrgeneral}
\eea
where 
\bea\label{deltaa}
\delta S_{a^2} =\frac{\rho}{16}({\rm Tr}(U_0+U_0))^2.
\eea
The superscript ``${\rm LO}$'' refers to leading order in the $\epsilon$-expansion and we
have introduced the dimensionless quantity 
\bea
\rho=F^2c_2a^2V.
\eea
Notice that if $\langle O^{\rm LO}_{1,{\rm ct}}(x)O^{\rm LO}_{2,{\rm
      ct}}(y)\delta S_{a^2} \rangle=\langle O^{\rm LO}_{1,{\rm
      ct}}(x)O^{\rm LO}_{2,{\rm ct}}(y)\rangle\langle\delta S_{a^2}
  \rangle$, i.e for disconnected insertions, the leading correction in
  Eq. \pref{Defcorrgeneral} vanishes. This happens for instance for left handed (V-A)
current correlators \protect\cite{Hernandez:2002ds}, and more general for 
correlators which do not get 
zero-mode contributions at leading order.

The epsilon expansion in continuum ChPT allows predictions for fixed
topological sectors \cite{Leutwyler:1992yt}. Since chiral symmetry is
explicitly broken in lattice QCD with Wilson fermions, an exact definition for
the topological charge does not exist at nonzero lattice spacing. For this
reason we will only give results for observables where the sum over all
topological sectors has been performed. 

\subsection{Preliminaries}

Calculations of mesonic 2-point functions in the $\epsilon$-regime have
been pioneered in Ref.\ \cite{Hansen:1990un}. The integration over the
non-constant modes $\xi(x)$ is done perturbatively as in ordinary
chiral perturbation theory in the p-regime. We summarize the
corresponding propagators and other useful properties in Appendix \ref{appeps}.

The 
integral over the constant mode $U_0$  has to be done exactly. In our
particular case with $N_f=2$ we encounter integrals of the type
\bea
\Tr{g(U_0)} &=& \frac{1}{Z_{0}} \int_{SU(2)} [dU_0] \,g(U_0)
\, e^{\frac{\mu}{2}{\rm Tr}(U_0+U_0^\dagger) }\,,\label{DefSU2int}
\eea
where $Z_0$ is the continuum partition function associated to the zero
modes, 

\bea\label{zeta0}
Z_{0} =\int_{SU(2)} [dU_0] e^{\frac{\mu}{2}{\rm Tr}(U_0+U_0^\dagger) }\,,
\eea
and $\mu$ denotes the standard combination
\bea
\mu &=& m\Sigma V\,.
\eea
Quite generally, the integral \pref{DefSU2int} leads, at least for the
types of $g$ we are considering, to expressions involving modified Bessel functions
$I_{n}(z)$ with integer index $n$. They satisfy numerous recursion
relations \cite{Grad} which allow us to express all integrals in terms of two
Bessel functions, which we choose to be $I_2$ and $I_1$. In Appendix \ref{appint} 
we collect various integrals that one encounters in calculating
the PP and AA correlator in the GSM$^*$ regime.

As an example let us consider the expectation value of the quantity
$\delta S_{a^2}$, defined in Eq. \pref{deltaa},
 that is part of the correction in \pref{Defcorrgeneral}.
By using the integrals given in the Appendix we obtain
\bea\label{EVa2} 
\langle{\delta S_{a^2}} \rangle&=& \rho \left(1 - \frac{3}{2\mu}\frac{I_2(2\mu)}{I_1(2\mu)}\right)\,.
\eea

\subsection{The PP correlator}
For the PP correlator we introduce the definition 
\bea
 \langle P^a(x) P^b(y)\rangle &=& \delta^{ab} C_{PP}(x-y)\,,
\eea
which takes into account
translation invariance and the trivial dependence on the flavor indices.
In the GSM$^*$ regime,  $C_{PP}(x-y)$ can be written through NLO as the sum
of the continuum correlator and a correction proportional to $a^2$,
\bea\label{cpp}
C_{PP}(x-y) &=& C_{PP, {\rm ct}}(x-y) + C_{{PP},a^2}(x-y)\,.
\eea
The continuum correlator for generic $N_f$ at NLO (which corresponds to $O(\epsilon^2)$)
is given by \cite{Hansen:1990un} (see also \cite{Damgaard:2001js}) 
\begin{equation}\label{ppct}
    C_{PP, {\rm ct}}(x-y)=C_P+\alpha_P\bar{G}(x-y),
\end{equation}
where $\bar{G}(x-y)$ is the finite-volume massless scalar propagator defined in Eq.\ \pref{gbar} and
\begin{eqnarray}
C_P & = & -\frac{\Sigma_{\rm eff}^2}{8(N_f^2-1)}\left[\langle{\rm
    Tr}[(U_0-U_0^\dagger)^2]\rangle_{\rm eff}-\frac{1}{N_f}
\langle[{\rm Tr}(U_0-U_0^\dagger)]^2\rangle_{\rm eff}   \right],\\
\alpha_P & = & \frac{\Sigma^2}{4F^2(N_f^2-1)}\Bigg[
2N_f^2-4-\frac{2}{N_f}\langle{\rm Tr}(U_0^2)+{\rm Tr}(U_0^{\dagger 2})\rangle\nonumber\\
&  & \hspace{1.5cm}+ \frac{2}{N_f^2}\langle{\rm Tr}(U_0){\rm Tr}(U_0^\dagger)\rangle+\left(\frac{N_f^2+1}{N_f^2}\right)\langle({\rm Tr}U_0)^2+({\rm Tr}U_0^\dagger)^2 \rangle\Bigg].
\end{eqnarray}
The expectation values with the subscript ``eff'' are defined like in
Eq.\ \pref{DefSU2int} with $\mu$ replaced by 
\begin{equation}
\mu_{\rm eff}=m\Sigma_{\rm eff}V,
\end{equation}
where $\Sigma_{\rm eff}$ is the quark condensate at one loop\cite{Hansen:1990un}
\begin{equation}
\Sigma_{\rm eff}=\Sigma\left(1+\frac{N_f^2-1}{N_f}\frac{1}{F^2}\frac{\beta_1}{\sqrt{V}}  \right).
\end{equation}
$\beta_1$ is a so-called \emph{shape factor} and is defined in Eq. \pref{beta1}.

For the particular case $N_f=2$, after the explicit computation of the
zero-mode integrals according to Appendix \ref{appint}, one gets
\begin{eqnarray}
C_P & = & \frac{\Sigma^2_{\rm eff}}{2\mu_{\rm eff}}\frac{I_2(2\mu_{\rm
    eff})}{I_1(2\mu_{\rm eff})}\,,\label{CP}\\
\alpha_P & = & \frac{\Sigma^2}{2F^2}\left[2-\frac{1}{\mu}\frac{I_2(2\mu)}{I_1(2\mu)}   \right].
\end{eqnarray}
For the leading lattice correction to the continuum result, as given in
Eq. \pref{Defcorrgeneral}, we find the $O(\epsilon^2)$ contribution
\begin{equation}
C_{{PP},a^2}=\rho\frac{\Sigma^2}{2}\Delta_{PP}, 
\end{equation}
with
\begin{equation}
\Delta_{ PP}=  \frac{5\mu I_1^2(2\mu)-10I_1(2\mu)I_2(2\mu) -3\mu I_2^2(2\mu)}{2\mu^3I_1^2(2\mu)}.
\end{equation}
Interestingly, the correction $\Delta_{PP}$ is finite in the
limit $\mu\rightarrow 0$, as is easily checked using the leading order Taylor
expansions for the Bessel functions \cite{Grad}, $I_1(2\mu) \sim \mu$ and
$I_2(2\mu) \sim\mu^2/2$. We are not aware of a rigorous argument that
$\Delta_{PP}$ has to be  regular at vanishing $\mu$, since this
correction ceases to be valid for small enough quark mass where one enters the
Aoki regime. A singularity at $\mu=0$ would have been a clear signal for this
breakdown of our calculation, however, this signal is not present in the
result, at least not at the order in the chiral expansion we are working here.

For the matching with numerical results obtained in lattice simulations we are interested in
the correlation function integrated over the spatial components,
\begin{equation}
    C_{PP}(t) = \int d^3\vec{x} \,C_{PP}(x-y)\Big|_{y=0}=C_{PP, {\rm ct}}(t)+ \frac{L^3\Sigma^2}{2}\rho \Delta_{ PP}\label{result2PP}\,,
\end{equation}
where
\bea
C_{PP, {\rm ct}}(t) &=&  \frac{L^3}{2}\frac{\Sigma_{\rm eff}^2}{\muref}\frac{I_2(2\muref)}{I_1(2\muref)}+ 
\frac{T\Sigma^2}{2F^2}h_1(t/T)\left[2-\frac{1}{\mu}\frac{I_2(2\mu)}{I_1(2\mu)}
\right]\label{ppcon}\,.
\eea
The time dependence is given by the parabolic function $h_1$ defined
in Eq.\ \pref{pg3}.

\subsection{The AA correlator}

The AA correlator is computed along the same lines. For simplicity we
consider the time-component correlator and we define 
\bea
 \langle A_{0}^a(x) A_{0}^b(y)\rangle&=& \delta^{ab}C_{AA}(x-y) \,.
\eea
Similarly to the PP correlator we split $C_{\rm AA}$ at NLO into a continuum part and a correction proportional to the lattice spacing,
\bea
C_{AA}(x-y) &=& C_{AA, {\rm ct}}(x-y) + C_{{AA},a^2}(x-y)\,.
\eea
The continuum contribution at $O(\epsilon^6)$ for $x\neq y$ and
generic $N_f$ has been calculated before  \cite{Hansen:1990un} (see also \cite{Damgaard:2002qe}) and is given by
\begin{equation}\label{aact}
    C_{AA, {\rm ct}}(x-y)=\alpha_A   \partial_{x_0}\partial_{y_0}\bar{G}(x-y)+\beta_A K_{00}(x-y)+\gamma_A \partial_{x_0}\partial_{y_0}H(x-y),
\end{equation}
where the following short hand notation has been introduced:
\begin{eqnarray}
\alpha_A & =& \frac{F^2}{2}\langle\mathcal{J}_0\rangle_{\rm eff}+ \frac{N_f}{2}\frac{\beta_1}{\sqrt{V}}\langle\mathcal{J}_0\rangle \,, \label{defalphaA}    \\
\beta_A & =&\frac{N_f}{2}\left(2-\langle\mathcal{J}_0\rangle\right)\,,\\
\gamma_A & = & \langle{\rm Tr}(U_0+U_0^\dagger)\rangle\frac{\mu}{N_f}.\label{defgammaA}
\end{eqnarray}
The functions $K_{\mu\nu}$ and $H$ are given in Eqs. \pref{defK00} and \pref{defH}. Moreover, we have introduced the quantity
\begin{equation}\label{defj0}
\mathcal{J}_0=\frac{1}{N_f^2-1}\left[{\rm Tr} U_0{\rm Tr}U_0^\dagger  +N_f^2 -2 \right].
\end{equation}
Like for the PP correlator, the  subscript ``eff'' refers to the
substitution $\mu\rightarrow \mu_{\rm eff}$ in the zero-mode integrals.
For the particular case we are considering, $N_f=2$, the results \pref{defalphaA} -- \pref{defgammaA} reduce to 
\begin{eqnarray}
\alpha_A & = & F^2\left[1-\frac{I_2(2\mu_{\rm eff})}{\mu_{\rm eff}I_1(2\mu_{\rm eff})}\right]+2\frac{\beta_1}{\sqrt{V}} \left[1-\frac{I_2(2\mu)}{\mu I_1(2\mu)}\right]\,, \label{alphaa}\\
\beta_A & = & \frac{2}{\mu}\frac{I_2(2\mu)}{I_1(2\mu)}\,,\\
\gamma_A & = & \frac{2\mu I_2(2\mu)}{I_1(2\mu)}.
\end{eqnarray}
In analogy to the PP correlator, the $O(a^2)$ contribution can be
computed according to Eq. \pref{Defcorrgeneral}, and we obtain
\begin{equation}
C_{{AA},a^2}(x-y)=\frac{F^2}{2}\partial_{x_0}\partial_{y_0}\bar{G}(x-y)\rho\Delta_{AA}\, ,
\end{equation}
with
\bea
\Delta_{AA}=\frac{-5\mu I_1^2(2\mu)+10I_1(2\mu)I_2(2\mu)+3\mu I_2(2\mu)^2}
{\mu^3I_1^2(2\mu) }=-2\Delta_{PP}\,.
\eea
Note that this correction affects only the coefficient $\alpha_A$ in
Eq.\ \pref{alphaa}, which will be modified by the presence of lattice artifacts.

The O($a^2$) correction is, up to a sign and a factor two, the same as the
correction for the PP correlator. As far as we can see there is no deeper
reason for this. It is simply a consequence of the fact that the zero-mode
integrals for the leading order continuum PP and AA correlator are very
similar, both lead to the same contribution involving the ratio $I_2(2\mu)/ \mu
I_1(2\mu)$. 

By integrating over the spatial coordinates and using the properties
listed in Appendix \ref{appeps}, we obtain for $t\neq 0$
\begin{equation}
    C_{AA}(t)=   \int d^3\vec{x}\,C_{AA}(x-y)|_{y=0}= C_{AA, {\rm ct}}(t)-\frac{F^2}{2T}\rho\Delta_{AA},
\end{equation}
where the continuum for $N_f=2$ explicitly reads
\begin{eqnarray}
    C_{AA,{\rm ct}}(t) &=& -\frac{1}{T}\alpha_A+\frac{T}{V}k_{00}\beta_A-\frac{T}{V}\gamma_Ah_1\left(\frac{t}{T}  \right)=\label{aacon}\\
                 & = & -\frac{F^2}{T}\left(1-\frac{I_2(2\muref)}{\muref I_1(2\muref)} \right) -\frac{2\beta_1}{T\sqrt{V}} \left(1-\frac{I_2(2\mu)}{\mu I_1(2\mu)} \right)+\nonumber\\
&+ &  \frac{2T}{V}k_{00}\frac{I_2(2\mu)}{\mu I_1(2\mu)}
-\frac{2T}{V}h_1(t/T) \frac{\mu I_2(2\mu)}{I_1(2\mu)}\,.\nonumber
\end{eqnarray}
Here $k_{00}$ is another shape factor defined in the appendix, Eq.\ \pref{defk00}. 

\subsection{The PCAC mass}
The correlators in the previous section are given as functions of $m$, the shifted mass. 
This is the mass parameter in the chiral Lagrangian and a priori not an
observable. Here we compute the PCAC mass, defined in \pref{DefmPCAC}, and use
it in the next section to replace $m$ with $m_{\rm PCAC}$. 
 
We have already calculated the denominator of  \pref{DefmPCAC}, and the numerator can be done analogously. Let us define
\bea
\langle \partial_{\mu}A^a_{\mu}(x) P^b(y)\rangle &=&\delta^{ab} C_{\partial A P}(x-y)\,.
\eea
To leading order in the epsilon expansion we find the result
\bea
C_{\partial A P,{\rm ct}}(x-y) &=&\frac{\Sigma}{V} \frac{I_2(2\mu)}{I_1(2\mu)}\,.
\eea
Dividing this by the leading order result of $2C_{PP,{\rm ct}}$ in Eq.\ \pref{ppct} we obtain
\bea
m_{\rm PCAC} &=& \frac{\mu}{\Sigma V}\,=\, m\,.
\eea
This is just the result of continuum ChPT, where it is not surprising
because the PCAC mass stems from the PCAC Ward identity. Note,
however, that both  numerator and denominator contain non-trivial
Bessel functions which cancel in the ratio. This cancellation will no longer happen with the lattice spacing corrections included, since the PCAC relation no longer holds.

The leading correction to the numerator in the GSM$^{\ast}$ regime is given by
\pref{Defcorrgeneral} with $O_1=\partial_{\mu}A^a(x)$ and $O_2=P^a(y)$ (no sum
over $a$). The computation is straightforward as the ones in the previous
sections and we find
\bea
C_{\partial A P}(x-y) &=&  \frac{\Sigma}{V}\frac{I_2(2\mu)}{I_1(2\mu)}\left[
  1 - \frac{3\rho}{2\mu^2} 
\left(2- \frac{\mu I_1(2\mu)}{I_2(2\mu)}+\frac{\mu I_2(2\mu)}{I_1(2\mu)}   \right)\right]\,.
\eea
Dividing by $2C_{PP}$ given in \pref{cpp} we obtain the leading $O(a^2)$
corrections to the PCAC mass:
\bea
m_{\rm PCAC} &=& m \left[ 1 + \rho \left(\frac{2}{\mu^2} - \frac{I_1(2\mu)}{\mu I_2(2\mu)} \right)\right]\label{PCACmassGSMast}\,.
\eea
The key observation here is that the PCAC mass is equal to $m$, up to a correction of O($ma^2V$), which is $\epsilon^2$ higher in the epsilon expansion in the GSM$^*$ regime. 

\subsection{The correlators as a function of the PCAC mass}
The final step we have to do is to replace $m$ by $m_{\rm PCAC}$ in the correlators. 
We first invert result \pref{PCACmassGSMast},
\bea\label{mbympcac}
\mu &=& \tmu \left[ 1 - \rho \left(\frac{2}{\tmu^2} - \frac{I_1(2\tmu)}{\tmu I_2(2\tmu)} \right) \right]\,,
\eea
where 
\bea
\tmu&=&m_{\rm PCAC}\Sigma V.
\eea
In the NLO contributions of the correlators  we can simply replace $m= m_{\rm PCAC}$, $\mu = \tmu$, since the corrections are higher than this order. In the LO term, however, we have to use the full expression \pref{mbympcac}, which gives rise to additional corrections proportional to $\rho$.

Eq.\ \pref{mbympcac} has to be inserted into the Bessel functions $I_n(2\mu)$. Since the correction proportional to $\rho$ is $\epsilon^2$ higher in the epsilon expansion we can Taylor-expand,
\bea\label{expI}
I_n(2\mu) &=& I_n(2\tmu) - 2\rho\tmu  I^{\prime}_n(2\tmu) \left(\frac{2}{\tmu^2} - \frac{I_1(2\tmu)}{\tmu I_2(2\tmu)} \right) +\ldots \,
\eea
and drop the higher order terms. 
The final results for the correlators can be brought into the form
\bea
C_{PP}(t) &=& C_{PP, {\rm ct}}(t)+ \frac{L^3\Sigma^2}{2}\rho \Delta_{a^2}\label{result2PPtilde}\,,\\
C_{AA}(t)&=& C_{AA,{\rm ct}}(t) + \frac{F^2}{T}\rho\Delta_{a^2} \,,\label{resultAAmod}
\eea
where the continuum correlators are as in \pref{ppcon} and \pref{aacon}, but with the replacements $\mu\rightarrow \tmu$ and $\muref\rightarrow \tmueff$, with
\bea
\tmueff &=& m_{\rm PCAC} \Sigma_{\rm eff} V\,.
\eea
The correction $\Delta_{a^2}$, which depends on $\tmu$, captures the lattice spacing artifacts and reads
\bea
\Delta_{a^2} & =& \frac{4\tmu^2I_1^3(2\tmu) -11\tmu I_1^2(2\tmu) I_2(2\tmu) +2(3-2\tmu^2)I_1(2\tmu)I_2^2(2\tmu) +5\tmu I_2^3(2\tmu)}{2\tmu^3 I_1^2(2\tmu)I_2(2\tmu)}\,.
\label{Deltaafinal}
\eea
Note that it is regular at $\tmu=0$.

Eqs.\ \pref{result2PPtilde} and \pref{resultAAmod} are our final results for
the GSM$^*$ regime. (In Appendix \ref{ocorr} we also give the corresponding
expression for the vector current correlator.) 
These results are remarkable and perhaps surprising in two ways:
(i) The $\tmu$ dependence of the O($a^2$) correction is identical  for both
correlation functions. (ii) Besides the continuum LECs $F$ and $\Sigma$ only
one more unknown LEC enters these expression, the parameter $c_2$. The second
feature is very advantageous in practice when our results are used to fit
numerical lattice data.

A different question is the actual size of the O($a^2$) correction, which is
directly proportional  to $c_2$. In the next section we try to give at least a
rough answer to this question.

\section{Numerical tests}
\subsection{General considerations}

For the pseudo scalar and axial vector correlators, the leading O($a^2$) correction in the GSM$^*$ regime is just a shift of the
constant part. The question is how big this correction is in
a typical $\epsilon$-regime simulation. As a measure for the correction we study
the ratio
\bea\label{ratio}
 R_{\rm XX}&=&\left| \frac{C_{\rm XX}(T/2) - C_{\rm XX, ct}(T/2)}{C_{\rm XX, ct}(T/2)}\right|\,,
\eea
i.e.\ the relative shift of the correlators at $T/2$. 
The main unknown here is the coefficient $c_2$. Even though it plays a decisive
role in the phase diagram of the theory \cite{Sharpe:1998xm}, it is difficult
to obtain in numerical simulations. So far only the ETM collaboration has
obtained an estimate from their simulations with a twisted mass term
\cite{Michael:2007vn,Urbach:2007rt}. The data for the pion mass splitting
together with the LO ChPT prediction gives the rough estimate $-2c_2 a^2\approx
(185 {\rm MeV})^2$ at  $a\approx 0.086$fm, which translates into $|c_2|
\approx (550 {\rm MeV})^4$. The error, however, is fairly large because of the
large statistical error in the determination of the neutral pion mass. In
addition, this value for $c_2$ was obtained with the tree-level Symanzik
improved gauge action and the standard Wilson fermion action, and any change in
this setup can and probably will lead to a different value for $c_2$.\footnote{An analysis
\cite{Aoki:2006js} of quenched twisted mass lattice data led to a value
$c_2\approx (300 {\rm MeV})^4$.} 
Nevertheless, for lack of a better estimate we use $|c_2| = (500 {\rm MeV})^4$ in the following.

For the other parameters we use $F=90$MeV, $a=0.08$fm and a hypercubic
lattice with $N_T=N_L=24$, which corresponds to a box size $L=1.92$fm. This
implies $\rho \approx 0.75$. Even though this is slightly large we may still
count this as O($\epsilon^2$) as it should in order to be in the GSM$^*$
regime. 

\begin{figure*}[t]
\begin{center}
 \includegraphics[width=0.5\textwidth,angle=-90]{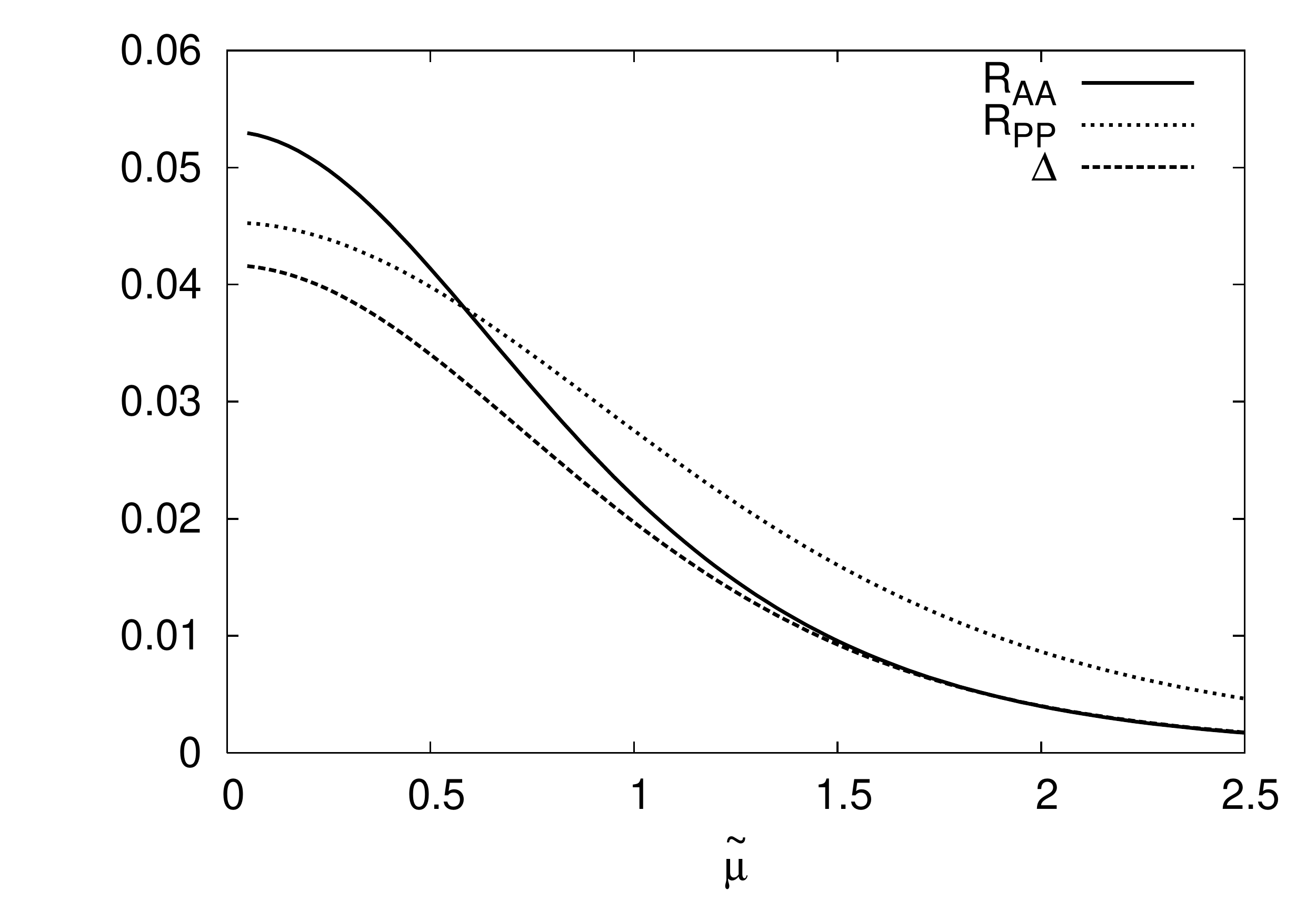}
 \end{center}
\caption{$ R_{PP}$ (dotted) and $ R_{AA}$ (solid) as a function of $\tmu$.
Both ratios are smaller than $3.5\%$ for $ \tmu \ge 0.75$. The dashed curve
represents $\Delta_{a^2}(\tmu)$.
 }
\label{fig:1}      
\end{figure*}

Figure \ref{fig:1} shows $ R_{PP}$ and $ R_{AA}$ for $\tmu$ values in the $\epsilon$-regime. 
For $\tmu = 1.0$ we find  $R_{ PP}=2.8\%$ and it decreases to less than
$1\%$ for $\tmu$ larger than 2. The correction is maximal (less than $5\%$) at
vanishing $\tmu$. However, for $\tmu\sim \rho$ we enter the Aoki regime and our
formulae are no longer valid. The values for $R_{AA}$ are very similar. For
instance, $R_{AA}=2.1\%$ at $\tmu = 1.0$.

Figure \ref{fig:1} also shows $\Delta_{a^2}(\tmu)$, which is, up to the factor
$\rho$,  the numerator in \pref{ratio}. It looks very similar to the ratios
itself, since the denominator in \pref{ratio} is of O($1$) and varies only
mildly for the $\tmu$ values considered her. So the correction to the
correlators is essentially $\Delta_{a^2}(\tmu)$, which happens to be of
the order of $10^{-2}$.

The main conclusion we can draw from this exercise is that for our choice
of parameters the O($a^2$)
corrections to the correlators are at the few percent level, a comfortably
small value. 

Using a bigger box improves the epsilon expansion since the expansion parameter
$1/(FL)^2$ is smaller. However, a bigger box also leads to larger $\rho$ values
and one easily enters the Aoki regime at moderately large volumes. For
instance, with $N_L=32$ and the other parameters unchanged we get $L=2.4$ fm
and $\rho\approx 2.4$, a value that is certainly not
O($\epsilon^2$).\footnote{It may seem counterintuitive at first that a change
in the volume may bring us into a different regime. However, increasing the
volume requires that we need to decrease the mass in order to stay in the
$\epsilon$-regime with fixed $\tmu$. Hence we have to decrease $a$ as well in
order to preserve the relative size between the mass and the lattice spacing
terms.}

\subsection{Reanalysis of recent lattice data}
In this section we investigate the impact of  $c_2$ on the extraction of the
continuum low energy constants $F$ and  $\Sigma$  from lattice data. The data
is taken from Refs.~\cite{Hasenfratz:2008ce,Hasenfratz:2008vi}. It is generated with $N_f=2$
flavors of dynamical improved  NHYP Wilson fermions \cite{Hasenfratz:2007rf} at
a fairly small quark mass. From there, a reweighting procedure allows to access
even smaller sea quark masses \cite{Hasenfratz:2008fg}. This procedure is exact
and does not introduce systematic uncertainties but allows to compute
correlators at very small quark masses at moderate cost. The lattice spacing is
$a\approx 0.115$fm from the measurement of the Sommer parameter $r_0$ taken to
be 0.49fm\cite{Sommer:1993ce}. We have two volumes available, one at $L=16a\approx1.84$fm and a
larger one with $L=24a\approx 2.8$fm. The former serves mainly as a cross check
whereas the latter has sufficient size for our NLO formulae to be applicable.
Some parameters of the simulation are given in Table~\ref{tab}.

\begin{table}
    \begin{tabular}{cclc}
	\ \ \	$L/a$ \ \ \ &\ \ \  $\kappa$\ \ \  &\ \ \  $a m_{\rm PCAC}$\ \ \  &\ \ \  $\mu$\ \ \ \\
	\hline
	24  &\ \ 0.128150\ \  &\ \  0.0019(4) \ \       & 2.1  \\
	    &\ \ 0.128125\ \  &\ \  0.0024(3) \ \      & 2.7  \\
	    &\ \ 0.128100\ \  &\ \  0.0030(3) \ \       & 3.4  \\
	    &\ \ 0.128050\ \  &\ \  0.0044(3) \ \      & 5.0  \\
	    \hline
	16  &\ \  0.128100\ \    &\ \  0.0028(11)\ \       & 0.7     \\
	    &\ \ 0.128050\ \    &\ \  0.0047(9)  \ \     & 1.1     \\
	    &\ \  0.128000\ \    &\ \ 0.0058(7)  \ \      & 1.4     \\
	    &\ \  0.127900\ \    &\ \ 0.0088(5)  \ \      & 2.1     \\
	    &\ \  0.127800\ \    &\ \ 0.0117(3)  \ \      &  2.9 
\end{tabular}
\caption{\label{tab}Parameters of the simulation. $L/a$ is the extend of the box, 
$\kappa$ the hopping parameters, the PCAC quark mass and 
an approximate values of $\mu=m\Sigma V$, where we use the central value of $\Sigma$.}
\end{table} 

\begin{figure}
    \begin{center}
\includegraphics[width=0.9\textwidth,angle=-90]{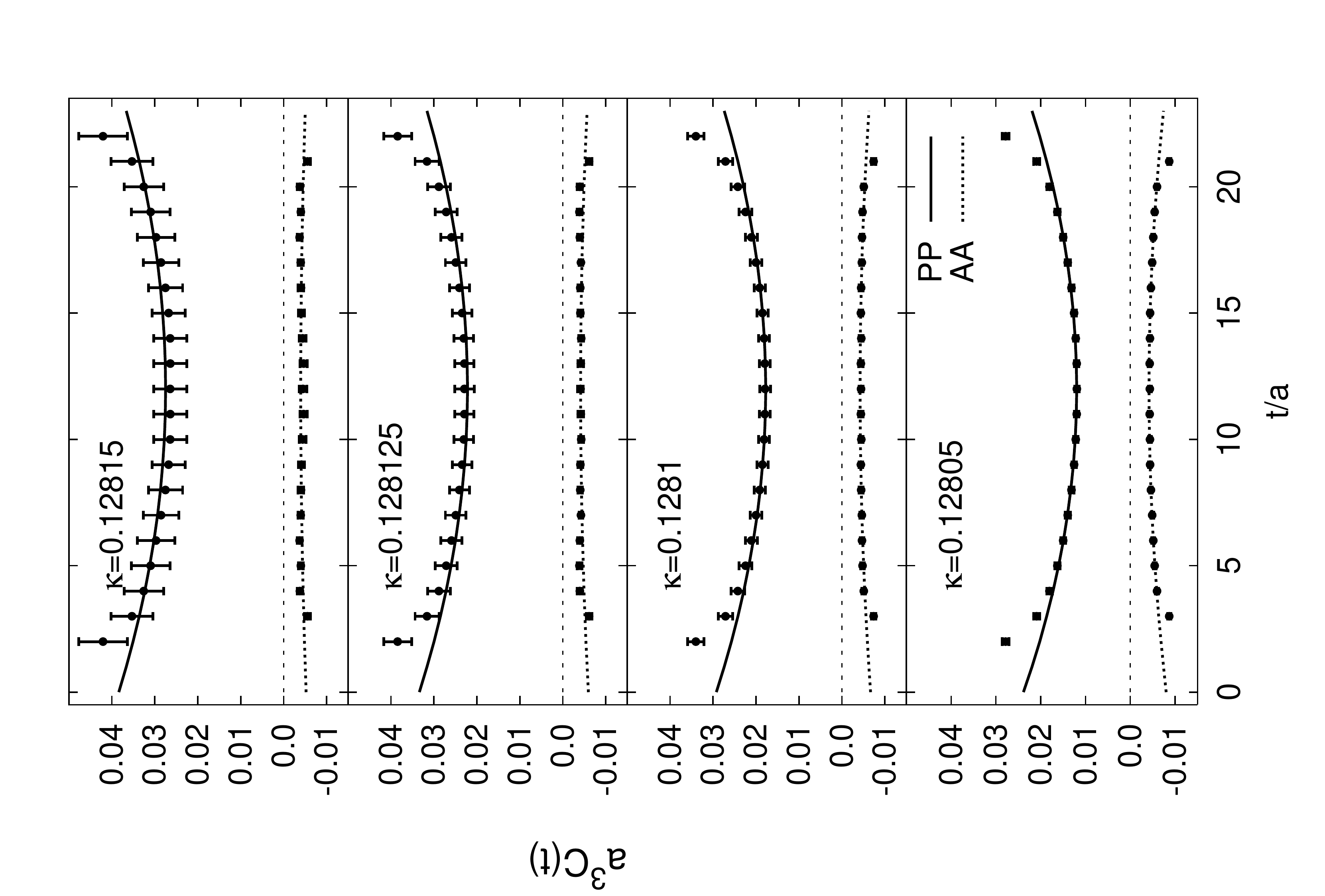}
    \end{center}
    \caption{\label{fig:fitlarge}Fit of the WChPT predictions to lattice data. All
    data points within the fit range of $t/a\in[6,18]$ for the four sea quark
    masses are included in the combined fit. The axial vector correlator is multiplied
    by a factor 50 for better visibility.}
\end{figure}

The theoretical formulae for the pseudo scalar and axial vector correlator
both have the form constant plus parabola. The coefficient $c_2$ only
contributes to the constant term in both cases. The curvature itself is rather
small at the parameter values simulated, in particular compared to the
statistical uncertainties, see Fig.~\ref{fig:fitlarge}. Therefore, at a fixed
mass, each of the two correlators effectively is a constant from which it is
difficult to constrain three parameters. As already discussed, the theory
predicts a particular and relatively strong $\tilde \mu$ dependence of the term
multiplied by  $c_2$. This gives a handle on the extraction of this
coefficient. Therefore we simultaneously fit the axial vector and pseudo scalar
correlators for all available quark masses. From a fit to $t\in[6,18]$ we get
$\Sigma^{1/3}=249(4)$MeV, $F=88(3)$MeV and $c_2=0.02(8)$GeV$^4$. The data,
along with the theoretical curves can be found in Fig.~\ref{fig:fitlarge}. Here
we used $Z_P^{\overline{\rm MS}}(2 {\rm GeV})=0.90(2)$ and $Z_A=0.99(2)$ from 
Ref.~\cite{Hasenfratz:2008ce}. The errors from the renormalization factors are
not included in the uncertainties of the LECs. The
value of $c_2$ is compatible with zero within errors and the one sigma band lies
within the range of reasonable values for a low energy constant. Since the data
points are highly correlated, we cannot give a good estimate for the quality of
the fit; we find $\chi^2/{\rm dof}=0.3(1)$ without the correlations taken into account.
We also remark that the results are independent of the fit range once $t_{\rm
min}/a >4$. Another concern are the relatively large values of $\tmu$. Therefore
we repeated the analysis leaving the $\tilde \mu\approx 5$ data out. We get
from the same fit range $\Sigma^{1/3}=250(4)$MeV, $F=87(3)$MeV and  $c_2=-0.01(8)$GeV$^4$. 
The differences to the previous values are well within the 
statistical uncertainties.
This is encouraging. Even with the additional constant the errors of the
continuum LECs are reasonably small. 

Is the value we find for $c_2$ large or not? To gauge
the impact of this term, we repeat the fit by setting $c_2=0$. The results are
virtually unchanged within errors: $\Sigma^{1/3}=249(4)$MeV, $F=88(3)$MeV. This
is very good news. The cut-off effects are so small that they do not impact the
extraction of the low energy constants beyond the level of the statistical
uncertainties.

As a cross check we repeated this analysis on the smaller volume, at the same
lattice spacing and  $L/a=16$. We obtain $\Sigma^{1/3}=257(4)$MeV, $F=83(2)$MeV 
and $c_2=0.06(14)$GeV$^4$. However, the (uncorrelated) $\chi^2/{\rm dof}=1.3$
might indicate that the NLO formulae are no longer applicable.  These results agree
with the findings of Ref.~\cite{Hasenfratz:2008ce}.

\section{Conclusions}
\label{Conclusions}

We have shown that the corrections due to the explicit chiral symmetry breaking
of Wilson fermions are highly suppressed. For typical quark masses these
corrections enter at either NNLO (GSM regime) or at NLO (GSM$^*$ regime). The
reason for this suppression can be traced back to the fact that the lattice
spacing corrections in the chiral effective action and the effective operators
are either quadratic in $a$ or they come with an additional power of either $m$
or $p^2$. There is no explicit term with a single power of $a$ only, since such
a term solely contributes to the additive mass renormalization which is
absorbed in the quark mass. Hence, the lattice spacing corrections are
suppressed in the chiral expansion, similar to the terms in the
Gasser-Leutwyler Lagrangian ${\cal L}_4$.

In the Aoki regime the modifications are more substantial, affecting the
correlators already at LO. The main complication in this regime are the zero
mode integrals, which are no longer the known Bessel functions. 

We tested our formulae against recent lattice data. We found that the additional
terms which come from the broken chiral symmetry have very little impact on the
extracted values of $F$ and $\Sigma$ whereas the low-energy constant associated
with the breaking is hard to determine precisely.

Our results derived here can be generalized in various ways, for example to the
case with a twisted mass term or to an arbitrary number of flavors. The details
of the calculation will change, but our various power countings can be carried
over with almost no modification. Perhaps most interesting from a practical
point of view is an extension along the lines of Ref.\
\cite{Bernardoni:2008ei}, where one considers a mixed setup with some quarks in
the $\epsilon$-regime and some others in the $p$-regime. 

However, the main conclusion one can draw is that the effects due to explicit
chiral symmetry breaking of Wilson fermions in the $\epsilon$-regime are less
severe than anticipated before. 
In view of the results of Ref.~\cite{Hasenfratz:2008ce} 
and the ones presented here, simulations with Wilson fermions
seem to be a viable alternative to the daunting task of dynamical simulations
with chiral fermions.

\section*{Note added} 

After this paper was completed we received a paper by A.~Shindler which  also deals with Wilson fermions in the $\epsilon$-regime and comes to similar conclusions \cite{AShindlereps}.
\section*{Acknowledgments} 

We would like to thank A.~Hasenfratz and P.~Hernandez for fruitful discussions and A.~Hasenfratz for 
reading the manuscript.

S.N. is supported by Marie Curie Fellowship MEIF-CT-2006-025673, and thanks the Physics Institute of the Humboldt University (Berlin) for hospitality during the preparation of this work.

This work is partially supported by EC Sixth Framework Program under the contract MRTN-CT-2006-035482 (FLAVIAnet),  by the Deutsche Forschungsgemeinschaft (SFB/TR 09) and the Ministerio de Ciencia e Innovaci\'on under Grant  No. FPA2007-60323 and by CPAN (Grant No. CSD2007-00042).

\begin{appendix}

\section{Some results for the epsilon regime}
\label{appeps}
In this Appendix we summarize formulae which are relevant for the computation of
correlation functions in the $\epsilon$-regime of chiral perturbation
theory. \\
Starting from the leading order continuum chiral Lagrangian of Eq.\ \pref{L2}
and by introducing the parametrization of Eq.\ \pref{eps_fac}, we can read off
the finite-volume scalar propagator for the nonzero modes:
\begin{equation}\label{gbar}
\bar{G}(x)=\frac{1}{V}\sum_{p\neq 0}\frac{e^{ipx}}{p^2},
\end{equation}
with 
$$
p=2\pi\left(\frac{n_0}{T},\frac{\vec{n}}{L}\right).
$$
The propagator $\bar{G}(x)$ satisfies the following properties:
\begin{eqnarray}
\int_V d^4x\, \bar{G}(x) &=&0\,,\label{pg1}\\
\partial_{\mu}\bar{G}(0)&=&0\,,\\
\Box\bar{G}(x)&=&-\delta(x)+\frac{1}{V}.\label{pg4}
\end{eqnarray}
UV divergencies, if present, are treated in dimensional regularization. \\
We define \cite{Hasenfratz:1989pk,Hansen:1990un}
\begin{eqnarray}
\bar{G}(0)& \equiv & -\frac{\beta_1}{\sqrt{V}}\,,\label{beta1}\\
T\frac{d}{dT}\bar{G}(0)& \equiv & \frac{T^2k_{00}}{V} \label{defk00},
\end{eqnarray}
where $\beta_1$ and $k_{00}$ are finite dimensionless \emph{shape coefficients} which
depend on the geometry of the box. They can be evaluated numerically: for
instance, for a symmetric box with $L=T$ one has $\beta_1=0.140461$ and
$k_{00}=\beta_1/2$ (see also \cite{Hernandez:2002ds}).

In order to obtain time correlators one has to perform integrals over
the spatial components of given functions of the propagators
$\bar{G}(x)$. In particular we define \cite{Hansen:1990un,Hasenfratz:1989pk} \footnote{In the original definition of \protect\cite{Hansen:1990un,Hasenfratz:1989pk}, $K_{\mu\nu}(x-y)$ contains also contact terms, which we do not consider in our computation since we are interested in the correlators for $x\neq y$.}
\begin{eqnarray}
K_{\mu\nu}(x-y) & = & \bar{G}(x-y)\partial_{x_\mu}\partial_{y_\nu}\bar{G}(x-y)-\partial_{x_\mu}\bar{G}(x-y)\partial_{y_\nu}\bar{G}(x-y)+\label{defK00}\\
                & + & \partial_{x_\mu}\partial_{y_\nu} H(x-y),   \nonumber\\
H(x-y)          & = &  -\frac{1}{V}\int_Vd^4z \bar{G}(x-z)\bar{G}(z-y). \label{defH}
\end{eqnarray}
The integrals that we need for this work are ($x_0=t$):
\begin{eqnarray}
\int d^3 \vec{x} \, \bar{G}(x-y)|_{y=0} & = & Th_1\left(\frac{t}{T}\right)=\frac{T}{2}\left[\left(\left|\frac{t}{T}\right|-\frac{1}{2}\right)^2-\frac{1}{12}   \right],  \label{pg3}\\
\int d^3 \vec{x}\, K_{00}(x-y)|_{y=0}  & = & \frac{T}{V}k_{00},\\
\partial_{x_0}\partial_{y_0}\int d^3\vec{x}\, H(x-y)|_{y=0} & = & -\frac{T}{V}h_1\left(\frac{t}{T}  \right).
\end{eqnarray} 

Finally, we recall the $SU(N_f)$ completeness relations which are used for
the computation of correlation functions.
Given the $SU(N_f)$ generators $T^a$, with $a=1,..,N_f^2-1$ and the convention
$$
{\rm Tr}[T^aT^b]=\frac{1}{2}\delta^{ab},
$$
one obtains
\begin{eqnarray}
{\rm Tr}(T^aAT^aB) & =& -\frac{1}{2N_f}{\rm Tr}(AB)+\frac{1}{2}{\rm Tr}(A){\rm Tr}(B),\label{com1}\\
{\rm Tr}(T^aA){\rm Tr}(T^aB) & =& -\frac{1}{2N_f}{\rm Tr}(A){\rm Tr}(B)+\frac{1}{2}{\rm Tr}(AB).\label{com2}
\end{eqnarray}
 
\section{SU(2) integrals}
\label{appint}
In the case $N_f=2$, the partition function related to the zero-mode integrals in
Eq.\ \pref{zeta0} is given by
\bea
Z_0=\int_{SU(2)}[dU_0]e^{\frac{\mu}{2}{\rm Tr}(U_0+U_0^\dagger)}= \frac{I_1(2\mu)}{\mu}\,,
\eea
where $I_n$ is the modified Bessel function of the first kind.
The normalization
\bea
\int_{SU(2)}[dU_0]=1
\eea
has been adopted.
Expectation values of arbitrary integer powers of ${\rm Tr} (U_0)$ can
be obtained by computing derivatives of $Z_0$. In particular, for this
work we need
\bea
\langle {\rm Tr}U_0\rangle & = &\frac{1}{Z_0}\frac{\partial
  Z_0}{\partial \mu}=2\frac{I_2(2\mu)}{I_1(2\mu)},\\
\langle \left({\rm Tr}U_0\right)^2\rangle & = &\frac{1}{Z_0}\frac{\partial^2
  Z_0}{\partial \mu^2}=4-\frac{6}{\mu}\frac{I_2(2\mu)}{I_1(2\mu)},\\
\langle \left({\rm Tr}U_0\right)^3\rangle & = &\frac{1}{Z_0}\frac{\partial^3
  Z_0}{\partial \mu^3}=-\frac{12}{\mu}+\frac{8(3+\mu^2)I_2(2\mu)}{\mu^2I_1(2\mu)},\\
  \langle \left({\rm Tr}U_0\right)^4\rangle & = &\frac{1}{Z_0}\frac{\partial^4
  Z_0}{\partial \mu^4}=16 +\frac{60}{\mu^2}-\frac{24(5+2\mu^2)I_2(2\mu)}{\mu^3I_1(2\mu)}.
\eea
Other integrals needed in this work can be related to the previous ones, for instance:
\bea
\langle  {\rm Tr}U_0^2\rangle & =& 2-\frac{3}{\mu} \langle {\rm Tr}U_0\rangle =2-\frac{6}{\mu}\frac{I_2(2\mu)}{I_1(2\mu)}\,,\\
\langle  {\rm Tr}U_0^2 \left({\rm Tr}U_0   \right)^2\rangle & =&
 -\frac{6}{\mu^3}\langle {\rm Tr}U_0\rangle+2\left(1+\frac{3}{\mu^2}\right)\langle \left({\rm Tr}U_0\right)^2\rangle-\frac{3}{\mu}\langle \left({\rm Tr}U_0\right)^3\rangle \nonumber\\
& = & \frac{4(15+2\mu^2)}{\mu^2}
-\frac{12(10+3\mu^2)I_2(2\mu)}{\mu^3I_1(2\mu)}.
\eea

\section{Other correlators}
\label{ocorr}

Here we summarize the GSM$^*$ results for some other correlators. 
We start with the correlation function of the time component of two
vector currents,
\bea
 \langle V_{0}^a(x) V_{0}^b(y)\rangle&=& \delta^{ab}C_{VV}(x-y) \,,
\eea
which we again split into a continuum part and a correction proportional
to the lattice spacing,
\bea
C_{VV}(x-y) &=& C_{VV, {\rm ct}}(x-y) + C_{{VV},a^2}(x-y)\,.
\eea
In our notation the leading order vector current in the chiral effective
theory reads
\bea
V_{\mu,{\rm ct}}^{a} & = & -i\frac{F^{2}}{2} {\rm Tr}\left({T^{a}(U^{\dagger}\partial_{\mu}U+U\partial_{\mu}U^{\dagger})}\right)\,.\label{Vectorcu
rrentct}
\eea
The continuum contribution at $O(\epsilon^6)$ for 
generic $N_f$ has been calculated before  by Hansen \cite{Hansen:1990un} (see also \cite{Damgaard:2002qe}). After integrating over the spatial coordinates one gets
\begin{equation}
C^{ab}_{VV}(t)=\delta^{ab}\left[-\frac{1}{T}\alpha_V+\frac{T}{V}k_{00}\beta_V            \right],
\end{equation}
with
\begin{eqnarray}
\alpha_V & = & \frac{F^2}{2}(2-\langle\mathcal{J}_0\rangle_{\rm eff})+\frac{N_f}{2}\frac{\beta_1}{\sqrt{V}}(2-\langle\mathcal{J}_0\rangle),\\
\beta_V & = & \frac{N_f}{2}\langle\mathcal{J}_0\rangle.
\end{eqnarray}
The function $\mathcal{J}_0$ has been defined in Eq. \ref{defj0}.
In particular, for $N_f=2$ the result explicitly reads
\begin{equation}
C^{ab}_{VV}(t)= -\frac{F^2}{T}\left(\frac{I_2(2\muref)}{\muref I_1(2\muref)} \right)
-\frac{2\beta_1}{T\sqrt{V}} \left(\frac{I_2(2\mu)}{\mu I_1(2\mu)} \right)
+\frac{2T}{V}k_{00}\left(1-\frac{1}{\mu}\frac{I_2(2\mu)}{I_1(2\mu)}  \right).
\end{equation}
The O$(a^2)$ correction in terms of the PCAC mass is given by
\bea
C_{{VV},a^2}(t)&=& - \frac{F^2}{T}\rho\Delta_{a^2}\,,
\eea
where $\Delta_{a^2}$ is defined in Eq.\ \pref{Deltaafinal}. Comparing
this with the result for the AA correlator in \pref{resultAAmod} we
observe that the lattice spacing corrections in these two correlators
are, up to a sign, identical.

With both the AA and the VV correlator at hand we can trivially obtain
the correlation functions of right- and left-handed currents. For
example, with $L^a_{\mu} = [V^a_{\mu} - A^a_{\mu}]/2$ we find
\bea
C_{LL}(t) &=& \frac{1}{4}\bigg( C_{VV}(t) + C_{AA}(t)\bigg)\,,
\eea
and the O($a^2$) corrections cancel in the sum on the right hand side,
i.e.
\bea
C_{LL,a^2}(t)& =& 0
\eea
while the continuum part is given by \cite{Hansen:1990un,Hernandez:2002ds}
\bea
C_{LL,ct}(t)&=&\frac{1}{4}\Bigg[-\frac{F^2}{T}-\frac{N_f}{T}\frac{\beta_1}{\sqrt{V}}+N_f\frac{T}{V}k_{00}
-\frac{T}{V}\langle{\rm Tr}(U_0+U_0^\dagger)\rangle\frac{\tilde\mu}{N_f}h_1\left(\frac{t}{T}  \right)\Bigg]\nonumber\\
&=&\frac{1}{2}\Bigg[-\frac{F^2}{2T}-\frac{1}{T}\frac{\beta_1}{\sqrt{V}}+\frac{T}{V}k_{00}
-\frac{T}{V}\frac{{\tilde\mu}I_2(2\tilde\mu)}{I_1(2\tilde\mu)}h_1\left(\frac{t}{T}  \right)\Bigg].
\eea
The same result can be obtained by a direct calculation of the
correlator, of course.

Finally, the scalar correlator
\bea
\langle S^a(x) S^b(y)\rangle&=& \delta^{ab}C_{SS}(x-y)
\eea
vanishes identically in the chiral effective theory for $N_f=2$, as one
can check either by explicit calculation or by using $G$-parity.

\end{appendix}

\end{document}